\documentclass[twocolumn,aps]{revtex4}
\usepackage{graphicx}

\newcommand{\nl}{\nonumber \\}

\newcommand{\be}{\begin{equation}}
\newcommand{\ee}{\end{equation}}
\newcommand{\bea}{\begin{eqnarray}}
\newcommand{\eea}{\end{eqnarray}}

\newcommand{\Eq}[1]{Eq.\,(\ref{#1})}

\newcommand{\la}{\langle}
\newcommand{\ra}{\rangle}
\newcommand{\dg}{\dagger}

\newcommand{\bfsigma}{\mbox{\boldmath $\sigma$}}
\newcommand{\mb}{\mbox}
\newcommand{\ti}{\tilde}

\begin{document}
\draft

\title{Non-adiabatic geometric quantum computation with trapped ions}

\author{Xin-Qi Li$^{1,2,3}$, Li-Xiang Cen$^{1}$, Guoxiang Huang$^{2}$, Lei Ma$^{2}$,
        and YiJing Yan$^{3}$}

\address{$^{1}$National Laboratory for Superlattices and Microstructures,
         Institute of Semiconductors,
         Chinese Academy of Sciences, P.~O.~Box 912, Beijing 100083, China \\
         $^{2}$Key Laboratory for Optical and Magnetic Resonance Spectroscopy,
         and Department of Physics, East China Normal University, Shanghai 200062, China \\
         $^{3}$Department of Chemistry, Hong Kong University of Science and Technology,
         Kowloon, Hong Kong }

\date{\today}
%\maketitle

\begin{abstract}
We propose a non-adiabatic scheme for geometric
quantum computation with trapped ions.
By making use of
the Aharonov-Anandan phase, the
proposed scheme not only preserves the globally
geometric nature in quantum computation,
but also provides the advantage of non-adiabaticity
that overcomes the problem of
slow evolution in the existing adiabatic schemes.
Moreover, the present scheme requires only
two atomic levels in each ion,
making it an appealing candidate for quantum computation.
\end{abstract}
%%\vspace{5ex}
\maketitle

%% \vspace{3ex} \pacs{PACS numbers: 03.67.Lx, 73.61.-r, 89.70.+c}

%% \leftline{\bf\it 0. Introduction}
%% \vspace{2ex}

Conventionally, the controllable operations
in quantum computation (QC) are achieved
on the basis of {\it dynamical} origins
of quantum state evolutions \cite{Div95}.
In recent years, for the purpose of being fault-tolerant to
certain types of computational errors,
there are considerable interest in exploiting the possibility of
implementing quantum computation by
{\it geometrical} means,    %% \cite{Ber84,Wil84},
which have been termed as {\it holonomic} quantum
computation \cite{Zan99,Pac00,Ell01}.
%%%%%%%%%%%%%
Depending on the degenerate property of the eigenspace of the governing
Hamiltonian, the holonomy can be
either a simple Abelian Berry phase factor \cite{Ber84}
or a general non-Abelian unitary transformation \cite{Wil84}.
It has been shown that the universal quantum
computation can be implemented in principle
by holonomies \cite{Zan99,Pac00,Ell01}.
Up to date, several experimental proposals have
been suggested for geometric quantum computation,
using such as the nuclear magnetic resonance \cite{Jon00},
superconducting nanocircuits \cite{Fal00,Cho01}, trapped ions \cite{Dua01,Pac01},
and nonlinear optics \cite{Pac00a}.

The principle of the aforementioned geometric QC is
rooted  in the {\it adiabatic} evolution of quantum system,
which may thus imply a slow computing speed.
The adiabatically slow evolution may also
challenge the sustainment of required coherence in QC.
Therefore, geometric QC based on {\it non-adiabatic}
evolution should be desirable.
Very recently, Wang and Keiji suggested to exploit
a non-adiabatic evolution to
realize geometric QC in NMR system and
superconductor nanocircuits \cite{Wan01}.
%%%%%%%%%%%%%%%%%
Indeed, geometric phase exists in non-adiabatic evolving
quantum systems, which is in fact the Aharonov-Anandan
(A-A) phase \cite{Aha87}.
Strictly speaking, the A-A phase depends on
certain dynamical quantities such as the
rotating angular speed of external (magnetic)
field or state vector \cite{Bul88,Fev92,Ni95}.
In this sense, the A-A phase differs from the
adiabatic Berry phase.
However, the dependence of the A-A phase on the angular speed
is through
the closed path loop
depicted by the ending point of state vector, and hence is
global in nature that largely retains
the geometric sense of A-A phase.
%is largely retained.
%rooted in its global nature.
Accordingly, quantum computation based on the non-adiabatic A-A phase can
be reasonably regarded as a kind of geometric QC.

  In this work we propose a scheme for non-adiabatic
geometric quantum computation with trapped ions.
Besides removing the drawback of the slow adiabatic evolution,
the proposed non-adiabatic
scheme holds additional merits as follows.
Firstly, there is no need to design the reverse evolving path to eliminate
dynamical phases that occur in
the Berry phase-based, adiabatic geometric QC  operations in
nondegenerated systems \cite{Jon00,Fal00}.
Secondly, in comparison to the existing fully
holonomic QC schemes \cite{Cho01,Dua01},
the present one does not involve the complicated construction of
the degenerate eigenspace of driving Hamiltonian.
The above two merits stem from the following observations.
In a non-adiabatic quantum evolution, the geometric A-A phase is in general
accompanied by a dynamical phase.
However, if the evolving path is such designed that along it
the state vector is
always perpendicular to the driving (magnetic) field,
the resulting phase factor after a non-trivial
cyclic evolution will be purely geometric.
This feature has been exploited by Suter {\it et al}
in their seminal experiment for demonstrating
the A-A phase \cite{Sut88}.
%%%%%%%%%%%%%%%%%%%%%%%%%%%%%%%%%%%%%%%%%%%%%
Finally, only two atomic levels of each ion are needed in our
scheme. This merit alone is attractive, since the originally
proposed ion-trap QC scheme required three levels \cite{Cir95} and
the recently proposed holonomic ion-trap QC required four levels
\cite{Dua01,Pac01}.
We notice that valuable efforts on improving
ion-trap QC protocol by using only two levels have been carried
out in dynamic schemes \cite{Mon97,Chi01,Mol99}.
%%%%%%%%%%%%
In particular, the technique proposed in Ref.\ \onlinecite{Mol99},
which effectively couples the electronic states of a pair of ions
by virtually exchanging phonons, is similar to our present one.
The major contribution of our work is to perform QC by geometric means.
Viewing that the work of Ref.\ \onlinecite{Mol99} has in fact extended the ion-trap QC scheme
to finite temperature, our geometric scheme may also hold to similar regime,
although in the following we would restrict our description at zero temperature limit.

%% {Div95},{Zan99,Pac00,Ell01}, {Ber84,Wil84} ,{Aha87},
%% {Jon00,Fal00,Pac00a,Dua01,Cho01}, {Cir95,Mon97}
%% \vspace{2ex} \leftline{\bf\it 1. Single Ion Hamiltonian}
%%\vspace{2ex}

{\it  Model Description}.
For quantum logic with trapped ions, we assume that
each ion has two relevant internal states $|0\ra$ and $|1\ra$ with energy
separation $\omega_0$,
and, as usual \cite{Cir95}, can be selectively addressed by lasers.
Consider, for instance, the $j$th ion being exposed to a traveling-wave laser
field ${\bf E}({\bf z})={\bf E}_0\mb{cos}({\bf k}\cdot{\bf z}-\omega_L t+\phi)$
with frequency $\omega_L$, wave vector ${\bf k}$, and phase $\phi$.
Here ${\bf z}=z_0\hat{\bf z}(a+a^{\dg})$ is the
center of mass coordinate of the ion
in terms of the phonon raising (lowering) operator $a^{\dg} (a)$
and zero-point spread $z_0\equiv (\hbar/2M\omega)^{1/2}$,
where $M$ is the total mass of the ion chain, and $\omega$ the phonon frequency.
The resulting Hamiltonian reads
\bea\label{Hj1}
{\cal H}^{(j)} = \frac{\omega_0}{2} \sigma_j^z
    + \ti{\omega}_j \left[ \sigma_j^+ e^{i\eta(a+a^{\dg})-i\omega_Lt+i\phi}
    + \mb{H.c.}  \right] .
\eea
Here, the atomic Pauli operators $\sigma^z=|1\ra \la 1|-|0\ra \la0|$,
$\sigma^+=|1\ra \la 0|$, and $\sigma^-=|0\ra \la 1|$, are introduced. In \Eq{Hj1}
$\ti{\omega}_j$ is the Rabi frequency, and $\eta=({\bf k}\cdot\hat{\bf z})z_0$ is the
Lamb-Dicke parameter, which accounts for the coupling strength between
internal and motional states.
%%%%%%%%%%%%%%
In our geometric scheme, the phonon plays no roles
in one-bit operation, but it does involve in
two-bit logic gate implementation as
it effectively couples two ions together.

%% \vspace{2ex}
%% \leftline{\bf\it 2. One-bit rotation via non-adiabatic A-A phase}
%% \vspace{2ex}

{\it One-Bit Rotation via Non-adiabatic A-A Phase}.
As just mentioned, in single bit logic operation,
we do not need the phonon-assisted dynamical flipping.
This can be the case under situations such as (i) at zero temperature
(no phonon excitation), if $\omega_L < \omega_0$;
and (ii) at finite temperature, but
$\omega_0 -\omega_L\neq n \omega$ ($n$ is an integer).
In the absence of phonon participation, \Eq{Hj1} can be re-expressed as
\bea\label{Hj2}
{\cal H}^{(j)}_R = {\bf \Omega}_j\cdot{\bfsigma}_j ,
\eea
in the rotating frame (with angular velocity $\omega_L\hat{\bf e}_z$).
Here notations have been introduced for
the effective magnetic field ${\bf \Omega}_j\equiv\{{\ti\omega}_j\mb{cos}\phi,
      {\ti\omega}_j\mb{sin}\phi,(\omega_0-\omega_L)/2 \}$,
and the vector Pauli operator ${\bfsigma}_j\equiv\{\sigma_j^x,\sigma_j^y,\sigma_j^z\}$.

%%%%%%%%%%%%%%%%%%%%%%%%%%%%%%%%%%%%%%%%%%%%%%%%%%%%%%%%%%%%%%%%%%%%
\begin{figure}\label{Fig1}
\begin{center}
\centerline{\includegraphics [scale=0.3] {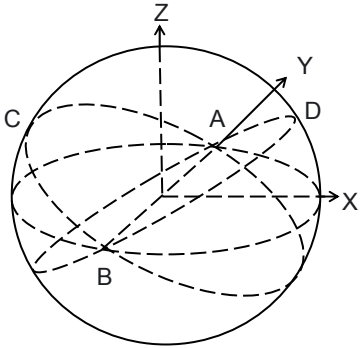}} \caption{
Schematic diagram for geometric rotation of state vector around
the {\it effective magnetic field}. Since the state is always
perpendicular to the field, there is no dynamical phase
accumulation during the evolution. }
\end{center}
\end{figure}
%%%%%%%%%%%%%%%%%%%%%%%%%%%%%%%%%%%%%%%%%%%%%%%%%%%%%%%%%%%%%%%%%%%%

To show how rotation between logic states $|0\ra$ and $|1\ra$ can be performed by geometric means,
we first consider the geometric evolution of the eigenstates of $\sigma^y$,
defined by $\sigma^y|\pm\ra=\pm|\pm\ra$, under appropriately designed laser sequences.
%%%%%%%%
Specifically, the state $|+\ra$, for instance, will complete a cyclic evolution
and acquire a non-adiabatic A-A phase under the following operations:
%%%%%%%%%%%%%%%%%%%
(i) Switching on a $\pi$-pulse with laser phase $\phi=0$, the state $|+\ra$
rotates around an effective magnetic field
${\bf \Omega}_1=\{ \ti{\omega}_j,0,(\omega_0-\omega_L)/2  \}$,
from $|+\ra$ in the $\hat{\bf e}_y$ direction to $|-\ra$ in the $-\hat{\bf e}_y$ direction
along the curve ACB on the Bloch sphere; see Fig.\ 1.
(ii) Suddenly changing the laser phase to $\phi=\pi$, after another $\pi$-pulse,
the state $|-\ra$ rotates back to $|+\ra$ around
${\bf \Omega}_2=\{ -\ti{\omega}_j,0,(\omega_0-\omega_L)/2  \}$ along
curve BDA on the Bloch sphere.
%%%%%%%%%%%%%%%%%%%
According to the A-A phase theory, after the above cyclic evolution, the state
$|+\ra$ will acquire a geometric phase $e^{i\gamma}$, with
$\gamma=4\arctan[2\ti{\omega}_j/(\omega_0-\omega_L)]$.
Note that during the above operation, the state vector
keeps always perpendicular
to the effective magnetic field, thus no dynamical
phase is accumulated in the evolution.
Similarly, the state $|-\ra$ will
acquire A-A phase $e^{-i\gamma}$ at the same time.

Now consider the evolution of logic states
$|0\ra=\frac{-i}{\sqrt{2}}(|+\ra-|-\ra)$,
and $|1\ra=\frac{1}{\sqrt{2}}(|+\ra+|-\ra)$.
After the above operations, they evolve to states
\bea\label{WF1}
|0\ra & \rightarrow &  \cos\gamma|0\ra+\sin\gamma|1\ra ,  \nl
|1\ra & \rightarrow &  \cos\gamma|1\ra-\sin\gamma|0\ra .
\eea
We see here the geometric A-A phase plays
a role of rotating the logic states.
Particularly,
complete state flipping
between $|0\ra$ and $|1\ra$ can take place at $\gamma=\pi/2$.
Note also that the possible value of $\gamma$
ranges from $0$ to $2\pi$, implying
the ability of arbitrary rotation between $|0\ra$ and $|1\ra$.

The state evolution of the performed qubit described by \Eq{WF1} is expressed in
the rotating frame with frequency $\omega_L$, in which other {\it free}
(not performed) qubits would have relative phase accumulations in the
non-resonant case of $\omega_L\neq\omega_0$.
Conventionally, a more convenient
choice is to express states in the interaction
picture with respect to ${\cal H}_0={\omega_0\over{2}}\sum_j \sigma_j^z$
(equivalently, a rotating frame with frequency $\omega_0$ around $z$-axis).
Accordingly, \Eq{WF1} can be recast in the interaction picture as
\bea\label{WF2}
|0\ra \rightarrow  e^{-i\omega_D\tau/2} \cos\gamma|0\ra
                 + e^{i\omega_D\tau/2}  \sin\gamma|1\ra ,  \nl
|1\ra \rightarrow  e^{i\omega_D\tau/2}  \cos\gamma|1\ra
                 -e^{-i\omega_D\tau/2}  \sin\gamma|0\ra ,
\eea
where $\omega_D=\omega_0-\omega_L$ and $\tau$ is the total operation time
on the performed qubit.

Interestingly, the above {\it state rotation (flipping)} is performed
by {\it non-resonant} pulses via geometrical means.
Now we show that by {\it resonant} pulses, a pure {\it phase shift gate}
of single qubit can be geometrically realized.
%%%%%%%%%%%%%%%%%%%%%
In resonant case, the laser-frequency-associated rotating frame coincides with
the interaction picture defined by ${\cal H}_0$, in which
the ($j$th) qubit Hamiltonian reads
$ {\cal H}^{(j)} = \ti{\omega}_j \left[ \sigma_j^+ e^{i\phi}
                 + \mb{H.c.}  \right]
                 = {\bf\Omega}_j\cdot{\bfsigma}_j  $,
with ${\bf\Omega}_j=\{\ti{\omega}_j\mb{cos}\phi,\ti{\omega}_j\mb{sin}\phi,0\}$.
We see that in the rotating frame the effective magnetic field constantly
has zero $z$-axis component for arbitrary laser phase $\phi$,
i.e., it lies in the $x$--$y$ plane.
%%%%%%%%%%%%%%%%%%%%%
To realize the single bit phase gate, we first switch on a $\pi$-pulse with
laser phase at certain value, say, $-\phi_0$.
The logic state $|0\ra$ and $|1\ra$ would rotate around the effective magnetic field
$\{\ti{\omega}_j\mb{cos}\phi_0,-\ti{\omega}_j\mb{sin}\phi_0,0\}$
to $|1\ra$ and $|0\ra$, respectively.
%%%%%%%%%
Then, suddenly changing the laser phase to $\phi_0$, after another $\pi$-pulse,
the logic states rotate around the effective magnetic field
$\{\ti{\omega}_j\mb{cos}\phi_0,\ti{\omega}_j\mb{sin}\phi_0,0\}$, and
return to the original states $|0\ra$ and $|1\ra$.
%%%%%%%%%%%%%%%%%%%%%%
Associated with this two step {\it cyclic} evolution, the logic states will
respectively acquire geometric A-A phases as
\bea\label{WF3}
|0\ra &\rightarrow& e^{i\ti{\gamma}} |0\ra  , \nl
|1\ra &\rightarrow& e^{-i\ti{\gamma}} |1\ra  ,
\eea
where $\ti{\gamma}=4\phi_0$.
%%%%%%%%%%%%%%%%%%%%
With the help of this phase shift operation, the additional phase factor in \Eq{WF2}
can be canceled out by properly choosing $\phi_0$.
More importantly, together with this phase shift gate,
the qubit state rotation \Eq{WF2} constitutes a complete logic set for arbitrary
single qubit operation.

%% \vspace{2ex}
%% \leftline{\bf\it 3. Two-bit gate via non-adiabatic A-A phase}
%% \vspace{2ex}

{\it Two-Bit Gate via Non-adiabatic A-A Phase.}
To complete the universal gate for quantum computation,
non-trivial two-bit gate
such as the controlled NOT, or equivalently, the conditional phase shift (CPS) gate,
would be required.
Below we show how the CPS gate can be implemented via geometric means.
%%%%%%%%%%%%%%%%%%%%%%%
Consider two qubits (e.g. the $j$th and $k$th ones) irradiated by two lasers
with frequencies $\omega_{L,1}$ and $\omega_{L,2}$, and phases $\phi_1$ and $\phi_2$.
By setting $\omega_{L,1}>\omega_0$ and $\omega_{L,2}<\omega_0$,
and correspondingly
denoting the detunings by $\delta_1=\omega_{L,1}-\omega_0$
and $\delta_2=\omega_0-\omega_{L,2}$,
the effective coupling between the two-bit
states $|00\ra$ and $|11\ra$ can be established
via virtually exchanging phonons, and
the resulting two-bit effective Hamiltonian reads \cite{Mol99}
%%%%%%%%%%%%
%%% If the detuning $\delta_1=\omega_{L,1}-\omega_0$ equals $\delta_2=\omega_0-\omega_{L,2}$,
%%% the conventional dynamic flipping between $|gg\ra$ and $|ee\ra$ would take place.
%%% In this paper, we are interested in the geometric realization of logic gates,
%%% thus we suggest $\delta_1\neq \delta_2$.
%%%%%%%%%%%%%
\bea\label{HT2}
\ti{{\cal H}}^{(j,k)}
   &=& \sum_{m=1}^{4} E_m |\ti{m}\ra \la \ti{m}| + g_{jk}
       \left[ e^{-i(\omega_{L,1}+\omega_{L,2})t}  \right.       \nl
   & & \left. \times e^{i(\phi_1+\phi_2)} \sigma_j^+\sigma_k^+ + \mb{H.c.}  \right] .
\eea
Here notations
$\{|\ti{1}\ra=|11\ra,|\ti{2}\ra=|00\ra,|\ti{3}\ra=|10\ra,|
 \ti{4}\ra=|01\ra \}$
are introduced for the two-bit computational basis states.
Up to the first-order expansion
of the Lamb-Dicke parameter $\eta$ in \Eq{Hj1},
the effective coupling strength can be obtained via
second-order perturbation theory as
$g_{jk}=g_jg_k(\frac{1}{\delta_1-\omega}-\frac{1}{\delta_2+\omega})$,
where $g_{j(k)}=\ti{\omega}_{j(k)}\eta$
is the one-phonon involved Rabi frequency
of single ion transition.
%%%%%%%%%%%%%%
The four basis-state energies $E_m$ ($m=1,\cdots,4$) contain also
the ac Stark shifts.
%which are not explicitly written out.
%%%%%%%%%%%%%%
The effective interaction couples only between
$|\ti{1}\ra$ and $|\ti{2}\ra$, but
leaves $|\ti{3}\ra$ and $|\ti{4}\ra$ inactive with respect
to the laser operation in study.
This effective two state dynamics (e.g. Rabi oscillations)
has been demonstrated in Ref.\ \onlinecite{Mol99} by numerical simulation
starting from the original Hamiltonian.
%%%%%%%%%
As a result, for two-bit operation,
one can focus on state evolution in the subspace $\{|\ti{1}\ra,|\ti{2}\ra\}$
in which the two-bit Hamiltonian (6) is recast to a similar
form as the single qubit \cite{Note1}
\bea\label{HR2}
\ti{{\cal H}}_{R}^{(j,k)} = \frac{\ti{\omega}_D}{2}\Sigma^z
     + g_{jk} \left( e^{i\Phi} \Sigma^+
     + e^{-i\Phi} \Sigma^- \right) ,
\eea
where $\Phi=\phi_1+\phi_2$,
and the {\it two-bit} Pauli matrices are introduced as
$\Sigma^z|\ti{1}(\ti{2})\ra=\pm|\ti{1}(\ti{2})\ra$,
$\Sigma^+|\ti{2}\ra=|\ti{1}\ra$, and $\Sigma^-|\ti{1}\ra=|\ti{2}\ra$.
Note also that the Hamiltonian (\ref{HR2}) has been expressed in the rotating frame
with respect to $\Sigma^z$ with the
rotation frequency $\omega_{L,1}+\omega_{L,2}$;
thus $\ti{\omega}_D=(E_1-E_2)-(\omega_{L,1}+\omega_{L,2})$.
Simple comparison of \Eq{HR2} with \Eq{Hj2} indicates that
an arbitrary rotation between
$|\ti{1}\ra$ and $|\ti{2}\ra$ can be performed geometrically
as that in the single bit case.

In the spirit of pulse-sequence operations
in dynamic scheme based on an XY spin model \cite{Los98,Ima99,Li02},
we shall in the following show that the CPS gate
can be implemented {\it geometrically} as
\bea\label{CPS}
{\cal U}_{\mbox{CPS}}
   = & & e^{i\pi/4}  e^{i\pi{\bf n}_j\cdot {\bfsigma}_j/3}
                       e^{i\pi{\bf n}_k\cdot {\bfsigma}_k/3}
             e^{-i\pi\sigma^x_k/2}    \nl
     & &\times
            U_{jk}(\pi/4)e^{-i\pi\sigma^y_j/2}
      U_{jk}(\pi/4)
            e^{-i\pi\sigma^x_j/2} .  % e^{-i\pi\sigma^x_k/2}  ,
\eea
Here the vector Pauli operator ${\bfsigma}=(\sigma^x,\sigma^y,\sigma^z)$,
unit vector ${\bf n}_j=(1,1,-1)/\sqrt{3}$,
and ${\bf n}_k=(1,-1,1)/\sqrt{3}$.
In the interaction picture with respect to the original free ion Hamiltonian,
the two-bit rotation operator  $U_{jk}(\alpha)$
is defined via $U_{jk}(\alpha)|01\ra = |01\ra$,
$U_{jk}(\alpha)|10\ra = |10\ra$, and
\bea\label{ROT0}
  U_{jk}(\alpha)|00\ra
       &=& \mb{cos}\left(\frac{\alpha}{2}\right)|00\ra
           +i\mb{sin}\left(\frac{\alpha}{2}\right)|11\ra ,  \nl
  U_{jk}(\alpha)|11\ra
       &=& \mb{cos}\left(\frac{\alpha}{2}\right)|11\ra
           +i\mb{sin}\left(\frac{\alpha}{2}\right)|00\ra .
\eea
%%%%%%%%%%%%%%%%%%%%%%%%%%
Since the arbitrary rotation and phase shift gate of single qubit have been built up,
the single bit operations in \Eq{CPS} can be readily implemented via geometric means
by properly combining the single bit logic elements.
Viewing the similarity between the two-bit reduced Hamiltonian (\ref{HR2}) in the subspace
$\{|00\ra,|11\ra\}$ and the single bit Hamiltonian (\ref{Hj2}),
the two-bit rotation of \Eq{ROT0} can be straightforwardly implemented
by the following two-step procedures:

(i)
In the {\it two-bit rotating frame} with frequency $\omega_{L,1}+\omega_{L,2}$
around $\Sigma^z$, performing cyclic evolution for the eigenstates of $\Sigma^y$
by controlling the laser phases $\Phi$ similarly as in the single bit case,
one can geometrically rotate the states $|00\ra$ and $|11\ra$.
Expressed in the interaction picture which also corresponds to
$\ti{\cal H}_0\equiv E_1|11\ra\la 11|+E_2|00\ra\la 00|$,
this operation can realize the following state transformation
\bea\label{ROT1}
|00\ra &\rightarrow&    e^{-i\ti{\omega}_D\tau/2}\cos\Gamma|00\ra
                    + e^{i\ti{\omega}_D\tau/2}\sin\Gamma|11\ra ,  \nl
|11\ra &\rightarrow&    e^{i\ti{\omega}_D\tau/2}\cos\Gamma|11\ra
                    - e^{-i\ti{\omega}_D\tau/2}\sin\Gamma|00\ra .
\eea
Here $\Gamma$ is the geometric A-A phase
determined by the evolution contour
of the two bit state vector,
and $\ti{\omega}_D\tau$ is the {\it detuning-induced} phase accumulation.

(ii) Tuning the laser frequencies in
resonance with the two ions \cite{Note2},
i.e., $\ti{\omega}_D=0$, a phase-shift gate associating
with $|00\ra$ and $|11\ra$ can be
implemented to cancel out the phase factors in \Eq{ROT1}.
In the resonance case
the $\omega_{L,1}+\omega_{L,2}$
rotating frame coincides with the interaction picture
of $\ti{\cal H}_0$.
As the one-bit resonant case, the effective magnetic field
corresponding to \Eq{HR2} now lies in the $x$--$y$
plane since $\ti{\omega}_D=0$.
By successively choosing two different values
of the laser phase $\Phi$, one can
perform two $\pi$ rotations on the states $|11\ra$ and $|00\ra$
around the effective magnetic fields,
and readily generate the A-A geometric phases,
$e^{-i\ti{\Gamma}}$ and $e^{i\ti{\Gamma}}$,
for states $|11\ra$ and $|00\ra$, respectively.
Now, after a phase-shift operation with
$\ti{\Gamma}=\ti{\omega}_D\tau/2-\pi/4$, \Eq{ROT1} becomes
\bea\label{ROT2}
|00\ra &\rightarrow&    e^{-i\pi/4} \left[ \mb{cos}\Gamma|00\ra
                     +  i \mb{sin}\Gamma|11\ra \right]   \nl
|11\ra &\rightarrow&    e^{i\pi/4} \left[ \mb{cos}\Gamma|11\ra
                     +  i \mb{sin}\Gamma|00\ra  \right] .
\eea
This is identical to \Eq{ROT0},
except for the additional global phases.
Obviously, these global phases
%Quite simply, the global phases in \Eq{ROT2}
will not appear if we first generate
a phase shift of $e^{i\pi/4}$ on $|00\ra$ and
$e^{-i\pi/4}$ on $|11\ra$ at the same time,
by the phase-shift gate just described, prior to the operation
of Eq.\ (\ref{ROT1}).

 We have thus realized the two-bit gate $U_{jk}$ as defined
in \Eq{ROT0}. Together with the arbitrary one-bit
operations (rotation and phase shift) described earlier,
we can now readily implement
the important CPS gate \Eq{CPS},
whose role is to
transform $|11\ra \rightarrow e^{i\pi} |11\ra $, while
to keep other computational two-bit basis states unchanged.

{\it Conclusion and Discussion.}
The proposed non-adiabatic geometric QC scheme based
on the A-A phases is expected to overcome
several drawbacks of the adiabatic schemes,   %% \cite{Jon00,Fal00,Cho01,Dua01},
namely, the slow evolution, need
of refocusing to eliminate the dynamical phases,    %% \cite{Jon00,Fal00},
and continuous control over
many fields to construct non-trivial loops in the parameter space.
%%%%%%%%%%%%%%%%%%%
Viewing that the trapped-ion is one of the
best exploited systems for quantum computation,
and that the proposed scheme requires a relatively simple atomic level
configuration,
we suggest, as a first step, to exploit it
as an interferometer for principle-proof
of the non-adiabatic geometric A-A phase discussed in this work.
We believe that the interference associated with the non-adiabatic
A-A phase can be readily demonstrated by
experiment in ion-trap systems.

  As a possible QC architecture, the elementary
operation steps in the proposed
non-adiabatic geometric scheme are comparable to its dynamic counterparts.
Specifically, the time scales for both the one-bit and two-bit
 geometric operations
are about the same as those in the dynamic operations.
%%%%%%%%%%%%%%%%%%%%%%%%%%%%%%%
Concerning the possible fault-tolerance,
in the adiabatic case, quantum logic is tolerant to
certain types of errors such as the field fluctuations
that preserve the loop area in parameter space, i.e., the Berry phase.
%%%%%%%%%%%%%%
Similarly, in the non-adiabatic case, the A-A phase
is of error tolerance to any fluctuations
around the state-evolving path that preserve the path loop area.
In principle, there exist
many possible driving field deviations that can
preserve the state path loop area.
However, the most natural and possible errors appear to
be random (but small) fluctuations
of the laser phase, frequency, and coupling strength to the atomic levels,
which equivalently result in fluctuations of the effective magnetic field.
The global A-A phase is expected to be largely
immune from this type of errors,
and, at the same time, the net dynamic phase accumulation
is approximately zero
due to the cancellation of the positive and negative contributions.

%% \section*{Acknowledgments}
\vspace{3ex}
{\it Acknowledgments.}
  Support from the Major State Basic Research Project
No.\ G001CB3095 of China, the Special Fund for "100 Person Project"
from Chinese Academy of Sciences,
the Youth Science and Technology "Qimingxing" Program of Shanghai (No.99QA14036),
and the Research Grants Council
of the Hong Kong Government is gratefully acknowledged.

%%%%%%%%%%%%%%%%%%%%%%%%%%%%%%%%%%%%%%%%%%%%%%%%%%%%%%%%%%%%%%%%%%%%%%%%%%

%%%%%%%%%%%%%%%%%%%%%%%%%%%%%%%%%%%%%%%%%%%%%%%%%%%%%%%%%%%%%%%%%%%%%%%%%%
\end{document}